# Gate-tuned quantum oscillations of topological surface states in β-Ag$_2$Te

*Azat Sulaev[1], Weiguang Zhu[2], Kie Leong Teo[3]\*, Lan Wang[1, 4]\**

*[1] School of Physical and Mathematical Sciences, Division of Physics and Applied Physics, Nanyang Technological University, Singapore 637371*

*[2]School of Electronics and Electrical Engineering, Nanyang Technological University, Singapore, 639789, Singapore*

*[3]Department of Electrical and Computer Engineering, National University of Singapore 117576*

*[4]Huazhong University of Science and Technology, School of Physics, Wuhan 430074, People's Republic of China*

We report the strong experimental evidence of the existence of topological surface states with large electric field tunability and mobility in β-Ag$_2$Te. Pronounced 2D SdH oscillations have been observed in β-Ag$_2$Te nanoplates. A Berry phase is determined to be near π using the Landau level fan diagram for a relatively wide nanoplate while the largest electric field ambipolar effect in topological insulator so far (~ 2500%) is observed in a narrow nanoplate. The π Berry phase and the evolution of quantum oscillations with gate voltage (V$_g$) in the nanoplates strongly indicate the presence of topological surface states in β-Ag$_2$Te. Moreover, the mobility of the narrow Ag$_2$Te nanoplate is ~ 3 × 10$^4$ cm$^2$s$^{-1}$V$^{-1}$ when the Fermi level is near the Dirac point. The realization of topological surface states with large electrical tunability and high mobility indicates that β-Ag$_2$Te is a promising topological insulator for fundamental studies.

Email: eleteokl@nus.edu.sg (Teo)

wanglan@ntu.edu.sg (Wang)

## I. INTRODUCTION

Topological insulator (TI) is a state of quantum matter characterized by $Z_2$ invariance [1-4]. It is an insulator in the bulk state but manifests conducting helical states at the boundary. The exotic boundary states of TIs are expected to form a playground of various topological quantum effects and show great potential in spintronics and quantum computation. Since the first experimental realization of TI in CdTe/HgTe/CdTe quantum well structures [5], an extensive effort has been put up in search for new TI systems. Till date, CdTe/HgTe/CdTe [5] and InAs/GaSb [6] have been experimentally confirmed to be two-dimensional TIs while strained HgTe and many Bi based compounds such as $Bi_xSb_{1-x}$, $Bi_2Se_3$ and $Bi_2Te_3$ have been realized as three-dimensional TIs [8-24]. To explore the possible application of TI in spintronics, ideal TI systems are expected to exhibit high mobility and gate electric field tunability. However, most of the TI materials show very small surface contribution due to defect-induced large bulk contribution and therefore electric field tunability cannot be realized. On the other hand, very few TIs that manifest electric field ambipolar effect such as $Bi_{2-x}Sb_xTe_{3-y}Se_y$ [26] and thin $Bi_2Se_3$ flakes [21, 24], show low mobility. On this basis, it is important to identify new TIs that show large gate electric field tunability and high surface mobility for the application of future TI based spintronic devices.

The $\beta$-$Ag_2Te$ is known for its unusual large and non-saturating quasi-linear magnetoresistance (MR) in the field range of $10$-$10^5$ Oe and temperature range of 5-300 K [27-29]. The origin of this unusual property has generated much debate since its discovery and may be associated with its 3D-TI nature. Theoretical calculation suggests that $\beta$-Ag$_2$Te is a TI with anisotropy and the unusual MR is largely originated from electrical transport on the topological surface states [30]. Recently, Aharonov-Bohm (AB) oscillations have been observed in the nanowire of $\beta$-Ag$_2$Te which indicates the existence of surface states [31, 32]. However, the topological nature of the surface state still remains to be confirmed.

In this paper, we report the further experimental evidence of the existence of topological surface states in $\beta$-Ag$_2$Te. Especially, the topological surface state of narrow Ag$_2$Te nanoplate exhibits the largest electric field ambipolar effect in TI so far ($\sim$ 2500%) and very high mobility $\sim$ 3 $\times$ 10$^4$ cm$^2$s$^{-1}$V$^{-1}$ near the Dirac point. We fabricated nanoplate devices as shown in the inset of Fig. 1(a) and Fig. 1(b). Both devices show pronounced two dimensional (2D) SdH oscillations. Using the measured resistivity $\rho_{xx}$ and Hall resistivity $\rho_{xy}$ of the nanoplate, the conductivity $\sigma_{xx}$ is calculated and the Berry phase is determined to be near $\pi$ by Landau Level (LL) fan diagram. This reveals that the SdH oscillation originates from the topological surface states. On the other hand, benefitted from the large ambipolar effect, the evolution of SdH oscillations with applied gate voltage ($V_g$) was investigated and also strongly indicates a Dirac cone composed surface state. Moreover, the mobility of the surface transport of the narrow $\beta$-Ag$_2$Te nanoplate is determined to be $\sim$ 3 $\times$ 10$^4$ cm$^2$s$^{-1}$V$^{-1}$ when the Fermi level is near the Dirac point. Finally, the MR variation with $V_g$ of the narrow nanoplate indicates a correlation effect between the bulk electrons and Dirac fermions.

## II. EXPERIMENTAL DETAILS

Using standard chemical vapor deposition (CVD) method, we have successfully obtained high quality β-Ag$_2$Te nanoplates. The details of growth method and characterization can be found in Ref. 31. The thickness of all the nanoplates varies from ~100 nm to ~300 nm, while the width varies from ~100 nm to ~20 μm. It is experimentally observed that the thickness of the β-Ag$_2$Te nanostructures does not change much with increasing growth time (from 10 mins to 3 hours). Longer growth time only results in the increase of width and length of the nanostructures. Standard photolithography technique was employed to pattern electrodes on the nanoplates. Cr/Au (5 nm/120 nm) contacts were deposited in a magnetron sputtering system with a base pressure of $1 \times 10^{-8}$ torr. Standard lock-in technique was utilized to perform four-terminal magnetoresistance measurements in two Quantum Design PPMS systems with 9 Tesla and 14 Tesla magnet respectively. All the measurements in this manuscript were carried out from -14 T to 14 T or from -9 T to 9 T, and then the $R_{xx}(H)$ and $R_{xy}(H)$ is calculated using formula

$$R_{xx}(H) = (R_{xx}(+H) + R_{xx}(-H))/2$$
$$R_{xy}(H) = (R_{xy}(+H) - R_{xy}(-H))/2$$

(1)

to eliminate the effect of the non-symmetric contacts.

## III. TEMPERATURE AND GATE VOLTAGE DEPENDENCE OF RESISTANCE

The typical transport behavior of two devices fabricated by a relatively wide nanoplate and a narrow nanoplate grown by CVD methods are shown in Fig. 1(a)-(d). It should be emphasized that similar results have been repeated in several devices fabricated using wide and narrow nanoplates. The wide β-Ag$_2$Te nanoplates tend to be slightly n-type. The inset of Fig. 1(a) and (b) show the fabricated devices using heavily p-doped Si with 300nm thick of $SiO_2$ dielectric layer. The thicknesses and widths for the two nanoplate devices are 120 nm (thickness), 5 μm (width) and 98 nm (thickness), 395 nm (width), respectively. As shown in Fig. 1(a), the resistance ($R_{xx}$) of the relatively wide nanoplate decreases slightly with temperature from $T$ = 300 K to 250 K, then increases about several times from 250 K to 30 K, and finally decreases to 10 K. The special temperature dependence of $R_{xx}$ of the nanoplate is due to the light impurity doping, which has been observed in many doped semiconductor systems[33]. The metallic behavior at low temperature of the nanoplate is attributed to the conduction in the impurity band, while the metallic behavior when T > 250 K can be explained by the thermal excitation of electrons from the Anderson localized states to extended states above the mobility edge. A better stoichiometry is achieved in the narrow nanoplate, and hence we observed much sharper increase of resistance with decreasing temperature as shown in Fig. 1(b). The gradually saturating behavior at low temperatures is due to the surface states. To probe the existence of surface states, we have performed the gate-tuned resistance measurement. As shown in Fig. 1(c), the $R_{xx}$ of the wide nanoplate only changes slightly with $V_g$ in view of electron doping. Although the $V_g$ dependence of $R_{xx}$ shows a peak near 0 V, the voltage dependence of Hall resistance ($R_{xy}$) and SdH oscillations indicates that electric field ambipolar gate effect cannot be realized in wide nanoplate devices due to charge doping. The resistance change with an applied gate voltage may originate from the variation of the density of states at the Fermi level in the bulk states when the Fermi level is slightly shifted by the gate electric field. We observed the phenomenon in several wide nanoplate samples. In contrast to wide nanoplates, the narrow nanoplates shows a huge ambipolar type electric field effect under a back $V_g$, which further confirms the high stoicheometry as indicated by the temperature dependence of resistivity curve. Figure 1(d) shows that the gate induced resistance change ($R_{peak}/R_{+50\ V}$) in a narrow β-Ag$_2$Te nanoplate is ∼ 2500%, a value which is much larger than that obtained in any other TIs. Here, the $R_{peak}$ and $R_{+50\ V}$ is the resistance at the peak of the ambipolar curve and $V_g$ = +50 V, respectively. The sharp transition at highest resistance point (the charge neutrality point) and the gradual change of resistance when the $V_g$ deviates from the charge neutrality point indicate that the ambipolar behavior could be due to topological surface states of Dirac cone. However, it should be emphasized that the ambipolar electric field effect can also be realized in semimetal or semiconductor with a very narrow band gap. In order to probe the topological nature of the surface state, we have performed two experiments. We first determine the Berry phase from the SdH oscillation of the conductivity of the wide nanoplate and then investigate the evolution of SdH oscillation with $V_g$ since the narrow nanoplate shows huge $V_g$ dependence of resistance of Dirac cone type.

## IV. BERRY PHASE OBTAINED FROM SdH OSCILLATIONs

Figure 2(a) shows the $R_{xy}$ of the wide nanoplate device under tilted magnetic field (*B*) as a function of the component of the magnetic field perpendicular to the sample surface ($B_\perp$). The angle (θ) is defined between the B field and sample surface. The $R_{xy}$ indicates an n-type carrier and a clear deviation from the linear relationship with *B* field. This demonstrates that there are more than one transport channel with different mobility values. At higher field, the $R_{xy}$ oscillates periodically in 1/B, which is the standard behavior of SdH oscillation. More importantly, we observe that the positions of maxima and minima do not change with $B_\perp$ field. This clearly indicates that the oscillations originate from 2D transport behavior. No oscillations are observed in $R_{xx}$ vs B curve (not shown here) when the applied magnetic field is parallel to the sample surface, which further supports the 2D transport and rules out the oscillation from bulk in this sample. The lack of the SdH oscillations from bulk electrons indicates low mobility of the bulk transport in this device.

A prominent property of Dirac fermions is that they carry the Berry phase of π. The observation of a π phase shift in SdH oscillation would clearly demonstrate that the 2D transport is indeed due to the topological surface transport[4]. SdH oscillations originate from successive emptying of Landau Levels (LL) with increasing magnetic field. The LL index n is related to the cross section area $S_F$ of the Fermi surface by

$$2\pi(n + \gamma) = S_F \frac{\hbar}{eB} \qquad (2)$$

where $\gamma = 0$ or 1/2 for topological trivial electrons and Dirac Fermions, respectively, *e* is the electron charge, h is the Planck constant ( $\hbar = \frac{h}{2\pi}$ ), and B is the magnetic flux density. As shown in Fig. 2(b), it is obvious that the SdH oscillation of $R_{xy}$ changes after the storage in vacuum and can be attributed to surface contribution. The conductivity is calculated using the formula $\sigma_{xx} = \rho_{xx}/(\rho_{xx}^2 + \rho_{xy}^2)$, thereafter the $\Delta\sigma_{xx}$ is obtained through a smooth back ground subtraction. The $\Delta\sigma_{xx}$ vs 1/B at 2 K (inset of Fig. 2(c)) shows oscillations of a single period which indicates only one transport channel in the surface states of the sample contributing to the oscillations. The LL fan diagram based on the oscillations of the conductivity is shown in Fig. 2(c), where its minima are identified to signify the integer n (indicated by arrows), while the half integers n+1/2 are assigned to the positions of maxima (indicated by arrows). The interception at 1/B = 0 should be at n = 0 (for the topological trivial surface states) or n = 1/2 (for the topological surface states). As shown in Fig. 2(c), a linear fit to the data gives interception of 0.54, which is close to the value of 0.5 expected for Dirac fermions. This result strongly supports that the SdH oscillation is indeed originated from the topological surface states.

Figure 2(d) displays the temperature dependence of SdH oscillations of $R_{xx}$. The inset shows the $\Delta R_{xx}$ obtained from $R_{xx}$ by subtracting a polynomial fit to the background. The amplitude of the SdH oscillations decreases with increasing temperature due to thermal agitation of electrons on the Landau levels. We have used the standard Lifshitz-Kosevich theory

$$\Delta R_{xx}(T,B) \propto \frac{\alpha T/\Delta E_N(B)}{\sinh[\alpha T/\Delta E_N(B)]} e^{-\alpha T_D/\Delta E_N(B)} \qquad (3)$$

to fit the temperature dependence of SdH oscillations，where $\Delta E_N = heB/2\pi m^*c$ is the energy gap between Nth and (N+1)th Landau Level, $T_D = h/4\pi^2 \tau k_B$ is the Dingle temperature, and $\alpha = 2\pi^2 k_B$. The $B$, $h$, $m^*$, $k_B$, and $c$ are the magnetic field, Planck constant, the effective mass of carriers, Boltzmann constant, and speed of light, respectively. The temperature dependence of $\Delta R/\Delta R(0\ K)$ for the 3rd Landau level is plotted in Fig. 2(e). The solid line is a fit to $\dfrac{\alpha T / \Delta E_N(B)}{\sinh[\alpha T / \Delta E_N(B)]}$. The $\Delta R$ (0 K) is the

$\Delta R$ at 0 K obtained from the fitting and $m*$ can be calculated using the fitted value of

$E_N$. We averaged the value obtained for different $B$ to get $m* = 0.12m_e$. From the

slope of the semi-log plot of $\Delta R B \sinh(\alpha T / \Delta E_N)$ vs $1/B$ at $T = 2$ K, the $T_D$ is determined to be 13.6 K and the carrier life-time is calculated to be 8.9 x $10^{-14}$ s. The mobility is then determined to be 1310 $cm^2 s^{-1} V^{-1}$.

## V. SdH OSCILLATIONS WITH GATE

The existence of Dirac cone composed topological surface states can be further investigated by examining the evolution of the SdH oscillation in the narrow nanoplate through back gating. From Eq. 2, we know that the period of $R_{xx}$ vs $1/B$ is determined by the cross section of Fermi surface as shown in Eq. 4,

$$\Delta \frac{1}{B} = \frac{2\pi e}{\hbar S_F}. \qquad (4)$$

Assuming a circular 2D Fermi surface or a spherical 3D Fermi surface, the value of

Fermi wave vector $k_F$ can be calculated using $S_F = \pi k_F^2$. Thereafter, the carrier

density of 2D surface states and 3D bulk states can be calculated by $n_{2D} = \dfrac{k_F^2}{2\pi}$ and

$n_{3D} = \dfrac{k_F^3}{3\pi^2}$, respectively. The dependence of $R_{xx}$ on $B$ field under various $V_g$ was

performed and the results are shown in Fig. 3(a)-(h). To analyze the SdH oscillations,

$\Delta R_{xx}$ is obtained from $R_{xx}$ by subtracting a polynomial fit to the background. The

$\Delta R_{xx}$ vs $1/B$ curves with various $V_g$s are shown in Fig. 4(a)-(f). The curves at -12 V

and -50 V do not show reasonable SdH oscillations, which may contribute to the very small Fermi surface and large noise, respectively. From Fig. 4, it is clearly evident that the period of SdH oscillation rises when $V_g$ varies from +50 V to -8 V. This demonstrates that the area of Fermi surface decreases when $V_g$ changes from +50 V to -8 V. With further increasing amplitude of $V_g$ to the negative direction, the magnitude of SdH oscillations gradually diminishes. The SdH oscillations can still be clearly observed and the period of the oscillations decreases when $V_g$ varies from -8 V to -25 V. This demonstrates that the area of Fermi surface increases in the procedure. This correlates well with the scenario that the area of the Fermi surface increases with the Fermi level shifting away from the Dirac point. In order to understand the band structure of $\beta$-Ag$_2$Te, we fit the $\Delta R_{xx}$ vs 1/B curve with the theoretical expression for SdH oscillations. The formula can be written as

$$\Delta R_{xx} = A \exp(-\pi / \mu B) \cos[2\pi(B_F / B + 1/2 + \beta)], \qquad (5)$$

where $B_F$ is the frequency of the SdH oscillation, A is the amplitude, $\mu$ is the mobility of carriers, and $\beta$ is the Berry phase[22, 23]. Equation 5 is in fact a zero temperature formula, which considers the effect of finite relaxation time but ignores the temperature effect[34]. As the accurate phase analysis should use the value of conductivity $\sigma_{xx} = \rho_{xx}/(\rho_{xx}^2 + \rho_{xy}^2)$, especially for systems like $\beta$-Ag$_2$Te that has similar values of $\rho_{xx}$ and $\rho_{xy}$, the $\beta$ value obtained in the fitting to $\Delta R_{xx}$ cannot provide the information of topological nature[4]. Here, we focus on two fitting parameters, the $B_F$ and $\mu$. The experimental data (red circles) and fitting curves (green lines) are shown in Fig. 4. Three transport channels are employed to fit the SdH oscillations at $V_g$ =

+50 V and $V_g$ = +25 V. One transport channel is utilized to fit the oscillations at other applied $V_g$s. The SdH oscillations at -8 V (Fig. 4(e)) show an abrupt increase at the large field, which probably originates from the deviation of background subtraction. A large deviation at large field can appear when the number of oscillations is small. The phase deviation of the fitting at large field at -25 V (Fig. 4(f)) may result from another transport channel or the large noise of the data. Although the fittings deviate in some features, our fitting do reveal the main features of the experimental data. The fitted $B_F$s and $\mu$s are shown in table 1. Based on the $B_F$ values, the values of $k_F$s are calculated as listed in the table. The $V_g$ dependence of $k_F$ is depicted in Fig 5(a), in which both the negative and positive value of $k_F$ is plotted as the red squares. For the SdH at +50 V and +25 V, we choose the largest $B_F$ to calculate the $k_F$. The green dashed line is drawn to clearly depict the cone structure. As previously discussed, the ambipolar effect can originate from the topological surface or bulk state of a semimetal. For the same Fermi surface area perpendicular to $B$ field, the bulk state has a much larger carrier density. Using the formulae as aforementioned, the calculated carrier densities at +50 V are $n_{3D}$ = $4.3 \times 10^{18}$ cm$^{-3}$ for the bulk state of semimetal and $n_{2D}$ = $4.0 \times 10^{12}$ cm$^{-2}$ for the Dirac cone composed surface state. At 0 V, the corresponding values for $n_{3D}$ and $n_{2D}$ are $3.0 \times 10^{17}$ cm$^{-3}$ and $6.8 \times 10^{11}$ cm$^{-2}$, respectively. We note that the charge tunability of a 300 nm SiO$_2$ dielectric layer is about $3.6 \times 10^{12}$ cm$^{-2}$/50 V. Considering the sample thickness of 98 nm, if the ambipolar behavior was due to the bulk state, the tuned charge would be $4.2 \times 10^{13}$ cm$^{-2}$ which is 11 times larger than the tuning ability of a SiO$_2$ dielectric layer. Thus, our results clearly rule out the possibility of bulk originated ambipolar behavior and provide further strong evidence of the existence of surface states of Dirac cone type. As for the other two channels showing SdH oscillations at $V_g$ = +50 V and +25 V, we speculate that they originate from the Rashba splitting induced topological trivial surface states, which have shown by the Altshuler-Aronov-Spivak oscillations in the β-Ag$_2$Te nanowire[31].

The $V_g$ dependence of the mobility is shown in Fig. 5(b), the carrier mobility is in the range between ~1 × 10$^3$ to ~2 × 10$^3$ cm$^2$s$^{-1}$V$^{-1}$ when the gate voltage varies from +50 V to +10 V. It increases to ~4 × 10$^3$ cm$^2$s$^{-1}$V$^{-1}$ at $V_g$ = 0 V and then get to ~3 × 10$^4$ cm$^2$s$^{-1}$V$^{-1}$ at $V_g$ = -8 V, which is a very large value compared with that in other topological insulators [11, 19-21, 35, 36]. Thereafter, the mobility decreases to ~2 × 10$^3$ cm$^2$s$^{-1}$V$^{-1}$ at $V_g$ = -25 V. The result indicates that the high mobility may be obtained when the Fermi level is tuned by an electric field to the Dirac point in topological insulators. It also agrees with the mobility value (1310 cm$^2$s$^{-1}$V$^{-1}$) obtained for the doped β-Ag$_2$Te nanoplate, in which the Fermi level is far away from the Dirac point. Using the width of the narrow nanoplate (395 nm) and the interval between the contact (7.2 μm), we can calculate the sheet resistance and then calculate the mobility using $\mu = \sigma/en$. Considering the devices have top and bottom surfaces, the calculated mobility at $V_g$ = 0 is ~ 5 × 10$^3$ cm$^2$s$^{-1}$V$^{-1}$, which also agrees well with the result obtained from SdH oscillation. For $V_g$s very near the charge neutrality point, we cannot get accurate $k_F$ due to the lacking of enough oscillations. The variation of MR with $V_g$ in the low field regime also shows an interesting feature. As depicted in Fig. 3(d), (e) and (f), the MR presents a relatively sharp increase in the low field regime (circled by green dash lines) at $V_g$ = 0 V, -8 V and -12 V, near the charge neutrality point and does not appear in the MR curves under other applied $V_g$. This may suggest that the π Berry phase induced weak anti-localization can be manifested when the $E_F$ is far away from the bulk band. In all, it is strongly indicated that β-Ag$_2$Te is a TI from the Berry phase of π obtained from LL fan diagram as shown in Fig. 2(c) and Dirac cone type evolution of SdH oscillations under $V_g$ as depicted in Fig. 3, 4 and 5.

## VI THE VOLTAGE DEPENDENCE OF MR

As aforementioned, the origin of the unusual MR of β-Ag$_2$Te is still under debate, which may originate from the bulk effect or surface effect[27-30]. To distinguish the bulk from the surface effect, high quality samples with tunable Fermi level are indispensable. It has not been realized prior to this work. Figure 6 shows the comparison of ambipolar effect and MR under applied $V_g$. The MR is defined as $(R(9\,T) - R(0\,T))/R(0\,T)$, which depicts a clear correlation with the $E_F$ position. The MR displays a lowest value of 78% at $V_g$ of the charge neutrality point. When the $V_g$ deviates from the charge neutrality point, the MR gradually increases in both negative and positive directions and reaches a maximum value at the voltages when the electric field ambipolar effect saturates. Thereafter, the resistance decreases with increasing $V_g$ magnitude in both negative and positive directions. Based on the relationship between the $E_F$ position and electric field ambipolar curve, we can clearly describe the MR behavior in β-Ag$_2$Te. At the charge neutrality point, the $E_F$ is located at the Dirac point and MR presents the lowest value. When the $E_F$ in the bulk band gap (but on the Dirac type surface states) is moved to the conduction band or valence band, the MR increases continuously until the $E_F$ touches the bulk band. Thereafter, any further manipulation to the bulk causes MR to decrease. Other than the bulk or surface origin of large quasi-linear MR, our experiment provides the possibility of controlling the interplay of bulk and topological surface state [37]. As shown in Fig. 4, the gated MR demonstrates that the largest MR appears near the top of the valence band and bottom of the conduction band, where the bulk electrons and Dirac Fermions can have a strong correlation. We still do not fully understand the correlation effect between the bulk electrons and Dirac Fermions, which deserves further theoretical and experimental investigations.

VII CONCLUSION

In conclusion, using LL fan diagram obtained from the oscillation of $\sigma_{xx}$ and the variation in the period of SdH oscillation with applied $V_g$, we provide strong evidence of the topological nature of surface states in β-Ag$_2$Te. The topological surface states of highly stoichiometric narrow β-Ag$_2$Te nanoplate exhibit the largest ambipolar effect in TI so far (~2500 %) and high mobility (~ $3 \times 10^4$ cm$^2$s$^{-1}$V$^{-1}$) near the Dirac point. This indicates that β-Ag$_2$Te is a very promising TI for fundamental studies. Moreover, the first report of $V_g$ dependence of MR in β-Ag$_2$Te suggests that the interplay between the bulk electrons and surface Dirac fermions has a large effect on MR for this material.

ACKNOWLEDGEMENT

We thank Ming Liang Tian for helpful discussions. This work was supported by the Ministry of Education of Singapore (Grant NO: MOE2010-T2-2-059), the Singapore A*STAR SERC 102 101 0019, and the National Science Foundation of China, Grant No. 61376127.

Tables

Table 1 The fitting parameter $B_F$ and $\mu$ for SdH oscillations, and the calculated $k_F$ from the fitted $B_F$ at various $V_g$s.

Figures

Fig. 1 (a) and (b) show the temperature dependence of resistance of β-Ag$_2$Te wide and narrow nanoplates, respectively. The insets show the corresponding device images. The red bars are 5 μm and 10 μm in (a) and (b), respectively. (c) and (d) show the $V_g$ dependence of resistance of β-Ag$_2$Te at 2 K.

Fig. 2 (a) Hall resistance ($R_{xy}$) as a function of the $B$ field measured at various tilted angles (θ); (b) $R_{xx}$ and $R_{xy}$ as functions of the $B$ field with θ = 90$^o$. The measurements were performed using the same sample for (a), which had been stored for three days in vacuum after the measurements in (a); (c) SdH fan diagram for measured 1/B with the filling factor n. The inset shows the Δσ$_{xx}$ vs. 1/B plot; (d) $R_{xx}$ vs. $B$ field curves at various temperatures. The inset shows the SdH oscillation after subtracting the background MR. (e) The temperature dependence of relative amplitude of SdH oscillation in ΔR$_{xx}$(B) for the 3$^{rd}$ LL. The solid line is a fit to eq. (1); (f) $\ln(\Delta R B \sinh(\alpha T / \Delta E_N))$ is plotted as a function of 1/B.

Fig. 3 $R_{xx}$ of the narrow plate as a function of $B$ field measured under various $V_g$, (a) +50 V, (b) +25 V, (c) +10 V, (d) 0 V, (e) -8 V, (f) -12 V, (g) -25 V and (h) -50 V.

Fig. 4 $\Delta R_{xx}$ as a function of 1/B field with various Vg, (a) +50 V, (b) +25 V, (c) +10 V, (d) 0 V, (e) -8 V, (f) -25 V. The red circles are the experimental data. The green lines are the fitting curves.

Fig. 5 (a) The variation of $k_F$ with $V_g$. The red squares are the calculated results based on the SdH oscillations. The green lines are plotted to depict the cone structure. (b) The $V_g$ dependence of the carrier mobility μ.

Fig. 6 The MR measured under various applied $V_g$. The $R_{xx}$ vs $V_g$ curve is plotted for comparison.

Table 1

| | $B_F$ (T) | $\mu$ ($\times 10^3$ cm$^2$s$^{-1}$V$^{-1}$) | $k_F$ ($\times 10^{-2}$ Å$^{-1}$) |
|---|---|---|---|
| +50 V | 83.5 ± 0.5 | 2.1 ± 0.2 | 5.04 ± 0.02 |
| | 70.5 ± 0.5 | 1.5 ± 0.2 | 4.63 ± 0.02 |
| | 45.6 ± 0.5 | 1.4 ± 0.2 | 3.72 ± 0.02 |
| +25 V | 52 ± 0.5 | 2 ± 0.2 | 3.98 ± 0.02 |
| | 44.1 ± 0.5 | 2 ± 0.2 | 3.67 ± 0.02 |
| | 10.6 ± 0.5 | 2.5 ± 0.2 | 1.80 ± 0.02 |
| +10 V | 24 ± 2 | 1 ± 0.2 | 2.70 ± 0.1 |
| 0 V | 14.1 ± 2 | 4 ± 0.5 | 2.07 ± 0.2 |
| -8 V | 10.2 ± 2.5 | 30 ± 5 | 1.75 ± 0.4 |
| -25 V | 25.1 ± 1 | 2 ± 0.3 | 2.76 ± 0.06 |

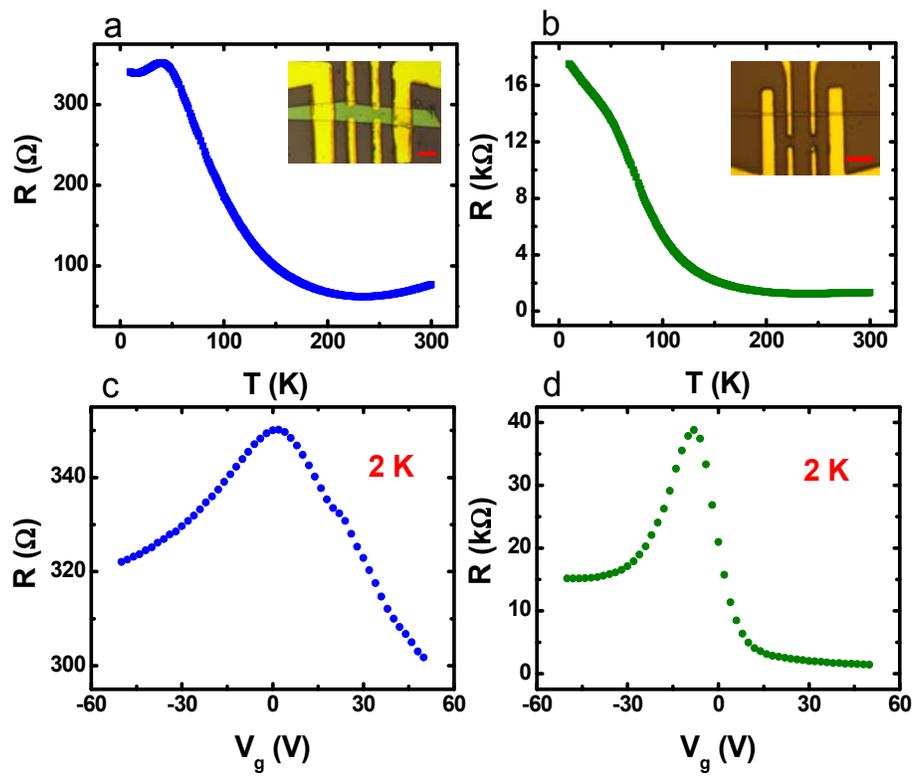

Fig. 1

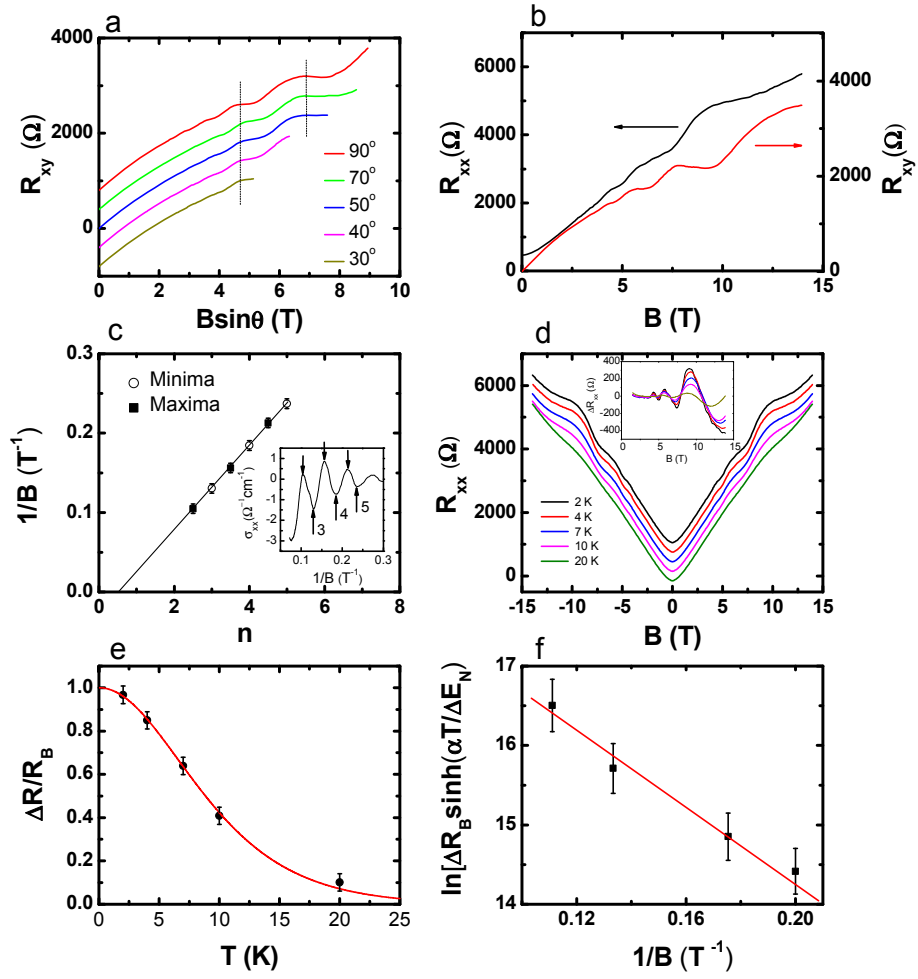

Fig. 2

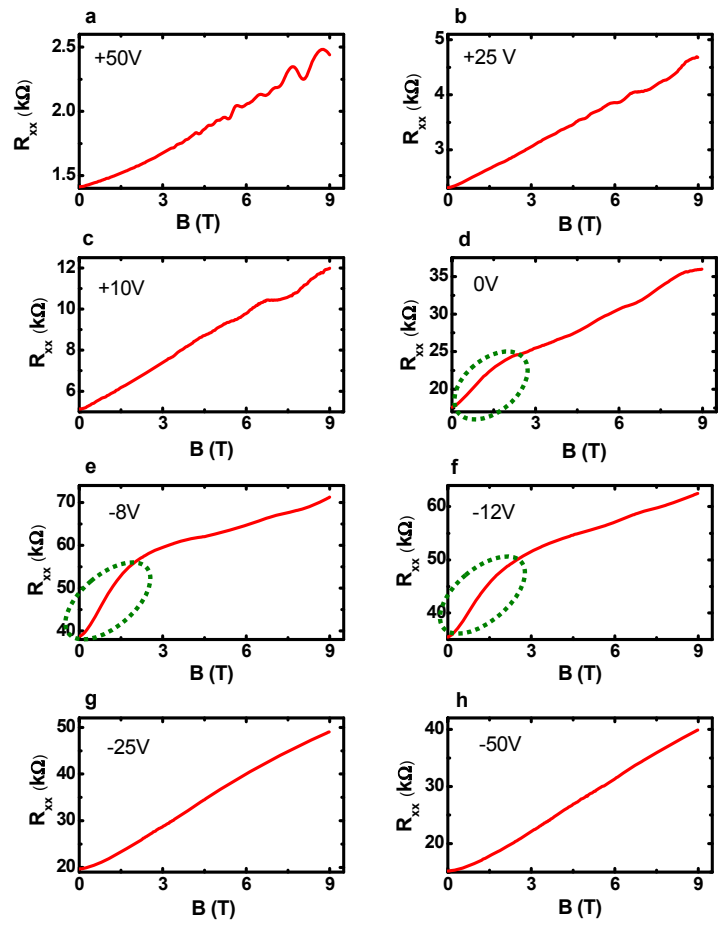

Fig. 3

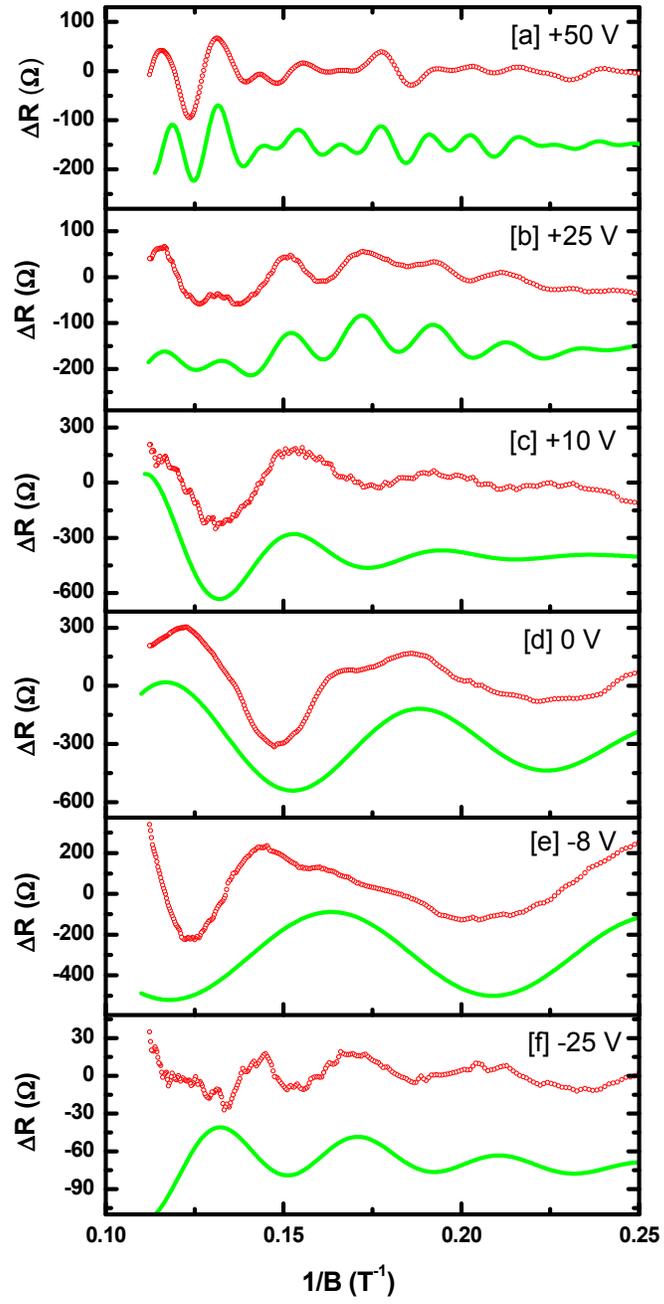

Fig. 4

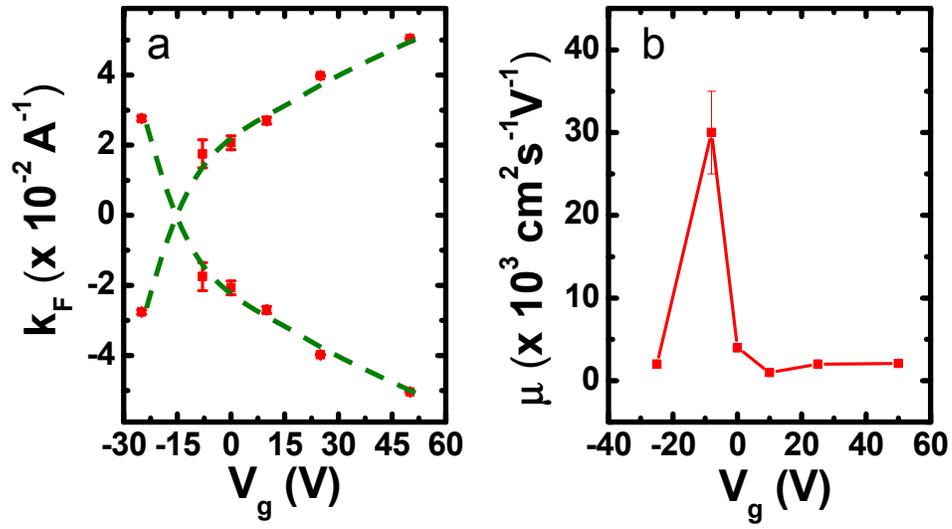

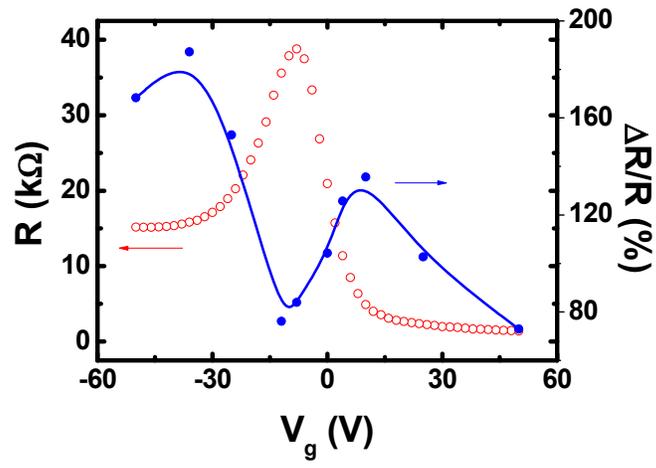

Fig. 6